\documentclass[prl,aps,twocolumn,showpacs,amsmath,amssymb]{revtex4}
\usepackage{epsf}

\begin{document}
\draft
\newcommand{\ve}[1]{\boldsymbol{#1}}

\title{Superconductivity Controlled by Polarization in
Field-Effect Devices of Confined Geometry}
\author{Natalia Pavlenko$^{1,2}$\cite{e-mail} and Franz Schwabl$^1$}
\address{$^1$Institut f\"{u}r Theoretische Physik T34, Physik-Department
der TU M\"{u}nchen, James-Franck-Strasse, D-85747 Garching
b.~M\"{u}nchen, Germany\\
$^2$Institute of Physics, University of Augsburg, 86135 Augsburg, Germany}

\begin{abstract}
We propose a concept for superconducting electric
field-effect devices based on superconducting films sandwiched between
ferroelectric layers. We provide theoretical calculations that indicate
how the field effect in these devices could be amplified, which
can be experimentally probed even at the current stage of film fabrication techniques.
\end{abstract}


\maketitle

The electric field effect in superconductors attracts considerable attention in
science and technology \cite{ahn}. The external field can modulate the charge
density and resistance, and control a reversible superconductor-insulator
switching behavior, which plays a key role in superconducting
field-effect transistors (SuFETs) based on dielectric or ferroelectric
polarisation. Especially in high-temperature superconducting oxides, where the
low carrier density $\sim 10^{21}$~cm$^3$ yields larger electric penetration
depths $\lambda_{TF}\sim 1$~nm, a switch of the ferroelectric polarization is
found to produce $\sim 10$~\% modulation in the carrier density $n$ at the
interface\cite{matthey}. The observed changes in $n$ and in the superfluid
density $n_S$ are shown to be the key factor responsible for a several K shift
in the superconducting transition temperature $T_c$ \cite{matthey,xi}, in
paricular in the underdoping region where $T_c$ is proportional to
$n_S(0)$ \cite{emery}. In the superconducting films of a thickness~$\gg
\lambda_{TF}$, the shift of $T_c$ due to the charge modulation in the interface
region of about 1~nm thickness is shunted. Thus, it is especially advantageous
for SuFETs to use ultrathin superconducting channels of a few nm thickness
\cite{logvenov,xi}. To make such SuFETs suitable for
technological applications, the achievement of $T_c$ shifts beyond the
range $5-10$~K is required. This needs the mechanisms of a possibly
stronger charge density modulation by the gate polarization in the range
10-30~$\mu$~C~cm$^{-2}$ which is currently achievable in ferroelectric oxides
Pb(Zr,Ti)O$_3$ (PZT) and (Ba,Sr)TiO$_3$ (BSTO).

\begin{figure}[htbp]
\epsfxsize=6.5cm \centerline{\epsffile{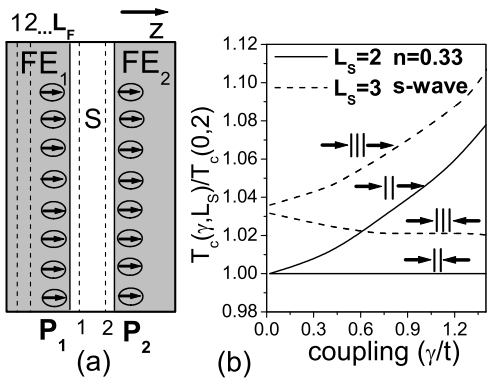}} \caption{(a) Scheme of
FE-S-FE sandwich with the uniform parallel spontaneous
polarization in FE-layers. (b) $T_c$ vs $\gamma$ in sandwiches containing $L_F=10$
FE-monolayers, where the cases of parallel and antiparallel gate polarization
are shown for a comparison. Here $V^0/t=-3.5$, $J_F/t=1$,
$t_{\perp}/t=0.1$, and the electronic band filling $n=0.3$. All the
temperatures are scaled by $T_c(\gamma=0,L_S=2)$, and we show here the case of
$s$-wave pairing in S-film.} \label{fig1}
\end{figure}

Recently, theoretical studies of superconducting-ferroelectric (S-FE)
multilayers \cite{pavlenko} have shown that the modulation of
charge in the ultrathin superconducting films sandwiched between the
ferroelectric layers is much stronger than in ferroelectric-superconductor
bilayers typically exploited for SuFETs. In a sandwich-like FE$_1$-S-FE$_2$
heterostructure (Fig.~\ref{fig1}(a)), the
polarization $P_2$ in the second gate FE$_2$ pushes an extra charge from the
interface S-FE$_2$ into the accumulation region at the interface
FE$_1$-S of the S-film. For high-$T_c$ cuprates as the most compatible
candidates for the sandwiches, the redistribution of
carriers between the interfaces driven by $P_1$ and $P_2$ could require S-films of 2-3 unit cell
thickness. In such films, the charge redistribution between CuO$_2$-planes
can occur via the interplanar tunneling of Cooper pairs which provides not only
the way for the charge modulation, but also enhances the local $T_c(loc)$ in
the accumulation region and thereby allows achievement of higher $T_c$ in the
entire S-film. For possible realizations of SuFETs based on
FE-S-FE sandwiches, the question that needs to be
addressed is how the superconducting properties can be controlled by the
voltages in FE-gates which is the subject of present studies.

Here we consider a superconducting film containing $L_S=2$ or $3$
infinite 2D-planes, as shown in Fig.~\ref{fig1}(a) for $L_S=2$. The
superconductivity in each plane is described by a BCS-like model with an
effective pairing potential $V^0$ (except that the energy cutoff is determined
by the electron bandwidth). These planes are weakly connected by the
interplanar Cooper pair tunneling with the tunneling energy
$t_{\perp}/t=0.05-0.1$ ($t \approx 0.1$~eV for high-T$_c$ cuprates is the
nearest neighbor hopping energy on a square lattice, which sets the energy
scale). The S-film is sandwiched between the FE-layers of a thickness given by
the number $L_F$ of unit cells in $z$-direction perpendicular to the
interfaces. In SuFETs, the charge redistribution in the S-film can be achieved
by reorienting the polarizations $P_1$ and $P_2$ in the layers FE$_1$
and FE$_2$ perpendicular to the interfaces by the gate electric field $E_g^1$
and $E_g^2$. Hence, we focus essentially on the two possible
orientations of ferroelectric dipoles (one of them is shown in Fig.~\ref{fig1}(a)),
representing them by two values $\pm 1/2$ of a pseudospin (dipole)
operator. To describe by this pseudospin formalism the nonzero
spontaneous polarization in each FE-layer due to ion displacements below the
Curie temperature, we employ an Ising model with the
dipole-dipole interaction energy $J_F$ taken into account in addition to the
interactions $-E_g^1\cdot P_1$ and $-E_g^2\cdot P_2$ with the gate fields
\cite{lines}. The screening of the polarization at the interfaces by the charge
in the S-film is described by the electrostatic charge- ferroelectric dipole
interaction $\gamma=\frac{e}{\Delta_{SF}^2}\cdot d_{FE}$. Here $d_{FE}$
is the magnitude of the dipole in each FE-unit cell and $\Delta_{SF}$ is the
distance between the nearest FE-unit cell and S-film. In our analysis, the
interface energy $\gamma$ ranges from zero (isolated S-film) to $\sim t$, which
should lie in a typical range of the charge-dipole interactions at the contacts
with ferroelectric BSTO(PZT)-layers of $\sim 100-300$~\AA~thickness where the
polarization $\stackrel{<}{\sim}25 \mu$~C~cm$^{-2}$ is suppressed due to strong depolarization
fields \cite{ghosez}. We study the system far below the Curie temperature,
treating the ferroelectric polarization in the mean-field approach \cite{pavlenko2,pavlenko3}.
Then the superconducting gaps and the polarization profiles are calculated
selfconsistently by the numerical minimization of the free energy of the system
\cite{pavlenko}.

We assume that the charge described by the electron concentration $n$ per
the unit cell in S-film, is already injected (resulting in the
superconducting state due to the effective attraction), and focus on the
question of {\it how the superconductivity can be affected by
the gate polarization}. When injected into the film, the charge
screens the interface polarization $P_1$ and $P_2$. The resulting
interplanar charge redistribution is described by the difference of the carrier
density in the boundary planes $\Delta n=n_1-n_{L_S} \approx
\frac{\gamma}{4t}(P_1+P_{2})$. The parallel polarization
($P_1=P_2=P>0$) leads to $\Delta n>0$, to the accumulation of
electrons in plane $i_S=1$ and their depletion in plane $i_S=L_S$. The
polarization $P_2$ in the second gate acts as an additional driving force
pushing the charge from the interface S-FE$_2$ to the plane $i_S=1$ and thus
providing stronger charge modulation in sandwiches.

The accumulation plane $i_S=1$ has higher local $T_c^1(loc)$ and leads to
increase of $T_c$ in the entire S-film due to the interplanar coupling
\cite{pavlenko}. For the hole carriers in the underdoped region of high-$T_c$
cuprates, the higher local $T_c(loc)$ is achieved in the electron depletion
(hole accumulation) plane $i_S=L_S$ and in the following discussion one should
consider the hole instead of electron accumulation. As compared to the isolated
film (case $\gamma=0$), the $P$-induced increase of $T_c$ is clearly seen in
Fig.~\ref{fig1}(b) for the parallel FE-gate polarizations ($P_1=P_2$).
The $T_c$ increase in this state does not
depend on the sign of $P_1$ and $P_2$, since their simultaneous switch
merely leads to the switch between the location of the accumulation and
depletion region without affecting the final $T_c$ \cite{pavlenko}. In contrast
to the state with parallel polarization, the antiparallel polarization in
the FE-gates ($P_1=-P_2>0$) results in a symmetric
electron accumulation at both interfaces so that $n_1=n_{L_S}$
($P_1=-P_2<0$ for hole accumulation). For the same amount of injected charge $n$, the
accumulation of carriers on both contacts does not provide significant charge
redistribution and leads to a decrease of the maximal achievable charge density
in the boundary planes $n_1(max)=n_{L_S}(max)=L_S n/2$, as compared to the
state with the parallel $P$ where the maximal possible accumulation density is
$n_1(max)=L_S n$. Consequently, the antiparallel polarization
results in lower $T_c$ (see Fig.~\ref{fig1}(b), where $P$ directed
towards the interfaces corresponds to electron accumulation). In particular,
for $L_S=2$, the antiparallel polarization does not give the redistribution of the injected charge
($n_1=n_2=n$) which results in the constant $T_c$ in Fig.~\ref{fig1}(b).

\begin{figure}[htbp]
\epsfxsize=8.0cm \centerline{\epsffile{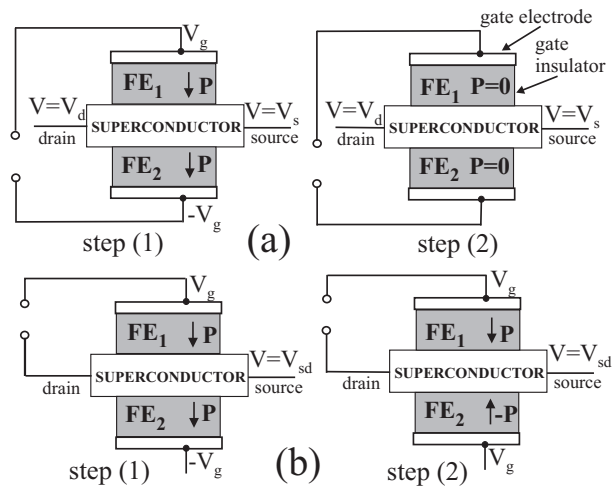}} \caption{Proposed schemes of
SuFET based on confined FE-S-FE geometry. In step(1) with parallel gate
polarization the SuFET is in the state with enhanced $T_c$, whereas step(2)
destroys (a) or decreases (b) the electron accumulation at FE$_1$-S interface and
thus switches SuFET into the state with low $T_c$.} \label{fig2}
\end{figure}

\begin{figure}[htbp]
\epsfxsize=8.5cm {\epsffile{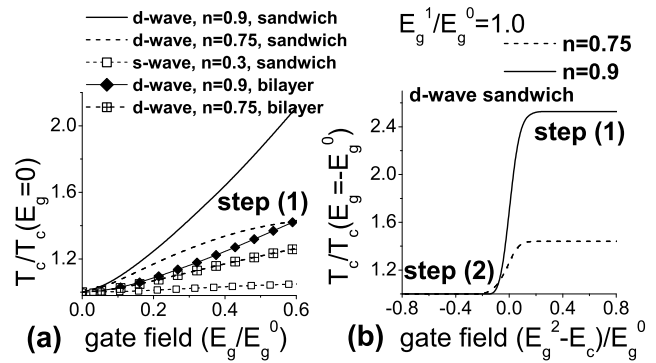}} \caption{Superconducting transition
temperature under variation of gate electric fields (a) for scheme shown in
Fig.~\ref{fig2}(a) where the gate fields $E_g^1=E_g^2=E_g$ set the
parallel polarization in both STO-gates of thickness $d=300$~nm with the
dielectric constant $\varepsilon=100$ and maximal achievable polarization
$P_{max}=5 \mu$~C~cm$^{-2}$; (b) for
scheme shown in Fig.~\ref{fig2}(b) where the FE$_1$-gate field is keeping fixed
whereas the FE$_2$-gate field is switched resulting in a reverse of the
FE$_2$-polarization at the coercive field $E_c$ in step (2). Here
$\gamma/t=0.5$ and FE-layers with a polarization $P_{S}=30 \mu$~C~cm$^{-2}$
are considered). In S-film for $d$-wave pairing, the planar band structure
$\varepsilon_{\ve{k}}=-2t(\cos k_x+\cos k_y)-4t_2 \cos k_x \cos k_y$ is chosen
with next-nearest-neighbor hopping $t_2/t=-0.4$, $t_{\perp}/t=0.05$ and
$V^0/t=-0.5$. All gate fields are scaled by a characteristic maximum
field $E_g^0\approx 10^6$~V/cm.} \label{fig3}
\end{figure}

Based on these advantages of sandwiches, we propose that the operation of an
SuFET containing an S-film {\it confined} between two FE-layers, can be
realized in two steps which are shown in two possible realizations in
Fig.~\ref{fig2}(a) and (b). Here, step(1) switches the SuFET to the
superconducting state with the enhanced $T_c$ caused by the parallel
polarization in the FE-gates. This can be achieved by applying the voltages
$V_g$ to the gate electrode FE$_1$ and $-V_g$ to FE$_2$. In the first
realization of step(1)(Fig.~\ref{fig2}(a)), the power supply simultaneously
moves the opposite charge to the gate electrodes. To reset the SuFET to the
state with lower $T_c$ (step(2)), one destroys the accumulation layer at the
FE$_1$-S contact, which is realized here by decreasing the gate voltages $V_g$
and tuning the gate polarization to zero. However, with the gates fabricated
from the PZT(BTO)-compounds, a nonzero spontaneous polarization $P_s \ne 0$ at
$V_g=0$ could result only in a partially removed accumulation region. Thus, in
this realization we propose to use the gates made with STO or BSTO with high Sr
content, so that $P_s\approx 0$ at temperatures close to $T_c$. The
corresponding modulation of $T_c$ is illustrated in Fig.~\ref{fig3}(a) for the
$s$- and $d$-wave pairing in the superconducting channel. As compared to the
FE-S-bilayers and $s$-symmetry channels, the remarkable shift of $T_c$ in
step(1) is obtained for the $d$-wave sandwiches, which strongly supports the
use of high-$T_c$ cuprates in the proposed SuFETs. For a more realistic
analysis of $d$-wave pairing, we choose the band structure with the
next-nearest-neighbor hopping which resembles the Fermi surface of
YBa$_2$Cu$_3$O$_{7-\delta}$. The electron density is taken near half-filling,
where the cases $n=0.9$ (hole density $x=1-n=0.1$) and $n=0.75$ ($x=0.25$)
should correspond to the under- and overdoped regions. There are a few
important points to note. First, for underdoping, the obtained increase of
$T_c$ is stronger than that for overdoping, which agrees well with the recent
observations of the strong field-effect in the underdoped region of the cuprate
phase diagram \cite{matthey}. Second, as the present studies are based
on a BCS-type model, the obtained here estimates for the $T_c$ increase in the
underdoped region consider only the mean-field boundary for the transition
temperature. To get more precise estimates for underdoping, the lowering of
$T_c$ by phase fluctuations \cite{emery} should be analyzed, which is the
object of future studies.

In the second realization (Fig.~\ref{fig2}(b)), the electric power sweeps the
charge from the FE$_1$-gate electrode and injects it into the
S-channel. Here, the FE$_2$-polarization is used to control the additional
enhancement of $T_c$ in the channel. Reversing the FE$_2$-polarization switches
the SuFET into the antiparallel polarization state with the lower $T_c$. The
modulation of $T_c$ by a change of the FE$_2$-gate field for the fixed
FE$_1$-gate field is shown in Fig.~\ref{fig3}(b), where the strong increase of
$T_c$ in a $d$-wave conducting channel is obtained when going from the step (2)
to the step (1) in Fig.~\ref{fig2}(b). Another possible modification of the second
realization could include the injection of additional charge from the FE$_2$-gate electrode
into the channel and thus achievement of higher $n$ (or $x$) in the state with antiparallel polarization.
Although the charge densities in both accumulation planes in this case
are comparable to those with the parallel $P$, the question of the
inter-planar charge modulation and thickness dependence of $T_c$ needs further
theoretical studies.

In conclusion, we have discussed the schemes of SuFETs based on
ferroelectric-superconducting sandwiches, where the $d$-wave channels
show great potential for strong modulation of $T_c$. However, for a successful
implementation, important theoretical and technological issues (related to the the growth of good
quality interfaces where the interface steps do not significantly affect the
charge redistribution in S-film) need to be solved.

This work was supported by the DFG Grants No.~SPP-1056, SFB-484 and
the BMBF Grant(13N6918A).

\end{document}